\documentclass[aps,pr,twocolumn,superscriptaddress,showpacs,floatfix]{revtex4}
\usepackage{amsfonts,amsmath}
\usepackage{graphicx}
\begin{document}

\title{Symmetry and dynamics universality of supermetal in quantum chaos}
\author{Ping Fang}
\affiliation{Department of Physics and Institute of Theoretical Physics
and Astrophysics, Xiamen University, Xiamen 361005, Fujian, China}
\author{Chushun Tian}
\email{ctian@mail.tsinghua.edu.cn}
\affiliation{Institute for Advanced Study, Tsinghua University,
Beijing 100084, China}
\author{Jiao Wang}
\email{phywangj@xmu.edu.cn}
\affiliation{Department of Physics and Institute of Theoretical Physics
and Astrophysics, Xiamen University, Xiamen 361005, Fujian, China}

\date{\today}

\begin{abstract}
Chaotic systems exhibit rich quantum dynamical behaviors ranging from
dynamical localization to normal diffusion to ballistic motion. Dynamical
localization and normal diffusion simulate electron motion in an impure
crystal with a vanishing and finite conductivity, i.e., an ``Anderson insulator''
and a ``metal'', respectively. Ballistic motion simulates a perfect crystal
with diverging conductivity, i.e., a ``supermetal''.We analytically find and 
numerically confirm that, for a large class of
chaotic systems, the metal-supermetal dynamics crossover occurs and is
universal, determined only by the system's symmetry. Furthermore, we
show that the universality of this dynamics crossover is identical to
that of eigenfunction and spectral fluctuations described by the random
matrix theory.
\end{abstract}

%05.45.Mt Quantum chaos; semiclassical methods
%64.70.Tg Quantum phase transitions
%72.15.Rn Localization effects (Anderson or weak localization)
%71.30.+h Metal-insulator transitions and other electronic transitions

\maketitle

\section{Introduction}

Chaotic systems exhibit a wealth of quantum phenomena.
A canonical example is the quantum kicked rotor~\cite{Casati1} (see
Refs.~\cite{Fishman10, Izrailev90, Chirikov79} for reviews)---a free
 rotating particle under the influence
of sequential time-periodic driving, the Hamiltonian of which reads
\begin{eqnarray}
\hat{H}=\frac{1}{2}(\tilde h\hat n)^2 + K\cos\hat{\theta} \sum_{m}\delta(t-m).
\label{eq:50}
\end{eqnarray}
Here the time $t$ is rescaled by the kicking period $\tau$, the Planck's
quantum $\tilde h=\hbar\tau/I$ with $I$ the particle's moment of inertia,
and the angular momentum $\hat n$ canonically conjugates to the angular
position $\theta$. The dimensionless amplitude of the kicking potential,
$K$, namely the so-called classical stochastic parameter, governs the
degree of the system's nonlinearity.
Despite this seemingly simple construction, tuning $\tilde h$ gives rise
to rich dynamical behaviors. For example, for (generic) irrational
values of $\tilde h/(4\pi)$ the rotor's kinetic energy saturates at long
times~\cite{Casati}, in sharp contrast to the linear energy growth in the
classical limit of vanishing $\tilde h$~\cite{Chirikov}. The former, the
so-called dynamical localization, simulates an Anderson insulator
\cite{Fishman} while the latter simulates a (normal) metal. Even more  striking,
$\tilde h$-driven phenomena occur to variants of the kicked  rotor
and they simulate a broad spectrum of condensed-matter systems (e.g.,
Refs.~\cite{KH1, KH2, KH3, KH4, KH5, Gong12, Altland11, Wang11, Gong15}).
Most recently, it has been found that for a spinful quasiperiodically
kicked rotor, the Planck's quantum can drive a sequential rotor Anderson
insulator-metal transition shown to be the mathematical equivalence of
the integer quantum Hall effect~\cite{Chen}.

A peculiar phenomenon common to various kicked rotor systems is the so-called
quantum resonance for rational values of $\tilde h/(4\pi)=p/q$, with $p$ and $q$ being
coprime natural numbers~\cite{QR}. For these values of $\tilde h$ the rotor's
kinetic energy grows quadratically at large times, which is much faster than
the metallic (linear) growth. Such a phase simulates a perfect crystal with
diverging electric conductivity corresponding to ballistic electron motion.
Therefore, it is dubbed ``supermetal''. Physically, for rational $\tilde h/(4\pi)$
the system exhibits translation symmetry in the angular momentum space, and
resonant transmission through the ensuing Bloch bands gives rise to the
supermetallic (quadratic) growth. The extreme sensitivity of the system's behavior
such as dynamical localization and quantum resonance to the number-theoretic
property of $\tilde h/(4\pi)$ is a prominent feature that distinguishes the
kicked rotor from genuine disordered systems (see Refs.~\cite{Tian, Altland11,
Casati, Dana, Wimberger, Izrailev90, Guarneri09, Fishman03a} for reviews of
the early and recent status of this subject). Over decades, this subject has
been central to theoretical and experimental studies of quantum chaos. Recently,
the supermetallic phase of the kicked rotor has been experimentally observed
in chemical systems and finds potential applications such as achieving
spin-selective rotational excitation of molecules~\cite{Milner12}.

So far, most studies have focused on the asymptotic supermetallic growth and
are mute to effects of chaoticity. In Ref.~\cite{Tian}, it is first found by
using the field theory that for the standard kicked rotor the chaoticity gives
rise to a universal crossover from metallic to supermetallic energy growth
which is insensitive to the system's details such as the values of $p$ and $q$ and
the strength of the kicking potential. Moreover, it has been found that this
universal dynamical behavior has a deep connection to the optical conductivity
of perfect crystals~\cite{Taniguchi93, Tian09} and the quantum walk in the
periodic multi-baker map~\cite{Dorfman1, Dorfman2, Dorfman}. A fundamental
question naturally arises: How robust is this dynamical behavior? Notably,
it is unclear whether such dynamics universality exists in general periodic
chaotic systems and is impervious to the symmetry. The present work aims at
a systematic study of these issues.

We study the supermetallic phase (rational $\tilde h/(4\pi)$) of a large
class of generalized kicked rotor systems. These systems have qualitatively
different dynamical behaviors (e.g., dynamical localization or delocalization)
when $\tilde h/(4\pi)$ is irrational. We analytically show and numerically
confirm that chaoticity leads to even richer universal behavior in the
crossover from metallic to supermetallic energy growth. We find that the
universal dynamics crossover behavior is very sensitive to the symmetry
of the Hamiltonian describing the free rotation of the particle, but not
to system's details. We show that the universal metal-supermetal dynamics
crossover can be attributed to the universality of fluctuations of the Bloch
wave functions and bands of the reduced quantum system in a unit cell. The
latter motivates us to conjecture that the universality class of supermetal
dynamics is identical to that of eigenfunction and spectral fluctuations
described by the random matrix theory (RMT)~\cite{RMT}.

The rest of the paper is organized as follows. In the next section we
introduce the models and discuss their symmetries. We present the analytical
results of the energy growth in Sec.~\ref{sec:fieldresults} and put them to
numerical test in Sec.~\ref{sec:simuresults}. In Sec.~\ref{sec:discussion}
we discuss the RMT foundation of the universality of supermetal dynamics.
We conclude in Sec.~\ref{sec:conclusion}. Some technical details are
presented in Appendix~\ref{sec:derivation}.

\section{\label{sec:Floquet} Model and symmetry}

In this work we will explore the dynamical behavior of a large class of
generalized quantum kicked rotors for rational values of $\tilde h/(4\pi)$.
In this section we first describe the models and discuss their symmetries.

\subsection{Generalized quantum kicked rotor}
\label{sec:KR}

The system to be considered below has a general Hamiltonian as follows,
\begin{eqnarray}
\hat{H}=H_0(\hat n) + K\cos\hat{\theta} \sum_{m}\delta(t-m).
\label{eq:13}
\end{eqnarray}
Here we assume the free rotation
Hamiltonian $H_0 (\hat n)$ to be an analytic function of $\hat n$ which can
be expressed as
\begin{equation}\label{eq:10}
H_0 (\hat n)=\sum_{k=0}^\infty c_k \hat n^k,
\end{equation}
where the numerical coefficients $c_k$ generally depend on $\tilde h$.
The angular momentum operator $\hat n$ canonically conjugates to the angular
operator $\hat \theta$ and has the eigenvalue spectrum $\{\tilde n\}, \tilde
n\in \mathbb{Z}$. When $H_0 (\hat n)=\frac{1}{2}\tilde h^2\hat n^2$ the system
is reduced to the standard quantum kicked rotor (\ref{eq:50}). The quantum
evolution can be expressed as a stroboscopic dynamics such that the wave
vector at integer time $t$, $| \psi(t) \rangle$, is given by
\begin{eqnarray}
| \psi(t) \rangle = \hat{U}^t | \psi(0)\rangle, \quad t\in \mathbb{Z},
\label{eq:6}
\end{eqnarray}
with the Floquet operator
\begin{eqnarray}
\hat{U}=e^{-\frac{i}{2\tilde h}H_0}e^{-\frac{iK}{\tilde h}\cos\hat\theta}
e^{-\frac{i}{2\tilde h}H_0}.
\label{eq:4}
\end{eqnarray}
As mentioned above, we shall focus on $\tilde h/(4\pi)=p/q$ throughout
this work.

\subsection{Symmetry}
\label{sec:symmetries}

First of all, we shall consider such $c_k$ that $H_0/\tilde h$ is shifted
by multiple $2\pi$ upon the translation operation
\begin{equation}\label{eq:11}
\hat T_q: \quad \hat n\rightarrow \hat n+q.
\end{equation}
As a result,
\begin{eqnarray}{\label{eq:12}}
[\hat{U},\hat{T}_q]=0,
\end{eqnarray}
implying that the system exhibits the translation symmetry in the
$\hat n$ space. This brings the rotor to the supermetallic phase.

Next, the system exhibits an effective time-reversal symmetry. To be
specific, we note that the Hamiltonian (\ref{eq:13}) is invariant under
the `time-reversal' operation
\begin{equation}\label{eq:14}
\hat T_c: \quad \hat n\rightarrow \hat n,\, \hat \theta\rightarrow -\hat
\theta, \, t\rightarrow -t.
\end{equation}
This has an important consequence, i.e.,
\begin{equation}\label{eq:15}
 \hat U^{\cal{T}}=\hat U,
\end{equation}
where the superscript `$\cal{T}$' stands for the transpose. This $\hat T_c$
symmetry brings the system to the orthogonal class in the RMT.

Finally, if $c_k=0$ for all odd $k$, then $H_0(\hat n)$ and thereby the
system bears an additional symmetry, i.e., $\hat U$ being invariant under
the operation
\begin{eqnarray}
\hat{T}_i: \quad \hat n \rightarrow -\hat n.
\label{eq:2}
\end{eqnarray}
Otherwise this symmetry is broken. This symmetry is dubbed the ``inversion
symmetry'' following solid-state physics \cite{Mermin}. Its effects on the
quantum resonance have not yet been explored~\cite{note1} and this is the
main subject of this work.

\section{\label{sec:fieldresults}Theory of supermetal dynamics}

In this section we present an analytic theory for dynamics of a generalized
quantum kicked rotor at resonance, i.e., the supermetallic phase. Particular
attentions are paid to effects of the inversion symmetry. Armed with this
theory, we explicitly calculate the rotor's `kinetic energy'~\cite{note2}
\begin{eqnarray}
E(t)\equiv\frac{1}{2} \langle \psi(t)|\hat{n}^2|\psi(t) \rangle.
\label{Etdf}
\end{eqnarray}
Throughout, to simplify technical discussions without loss of generality,
we assume that the initial state is an unperturbed eigenstate with zero
angular momentum, i.e., $|\psi(0)\rangle =|0 \rangle$.

\subsection{Reduced system}
\label{sec:reduction}

First of all, the quantum dynamics (\ref{eq:6}) in the angular momentum
space is determined by the quantum amplitude, $\langle \tilde n|\hat U^t|
\tilde n'\rangle$. It is easy to show that the time Fourier transformations
of this amplitude and its complex conjugate, denoted as $\langle \tilde
n|(1-e^{i\omega_+}\hat U)^{-1}|\tilde n'\rangle$ and $\langle \tilde n'|
(1-e^{-i\omega_-}\hat U^\dagger)^{-1}|\tilde n\rangle$, respectively,
give
\begin{eqnarray}
    E(t)&=&\frac{1}{2}\sum_{\tilde n} \tilde n^2 \int \frac{d\omega}{2\pi}
    e^{-i\omega t}\label{eq:21}\\
    &\times& \langle\! \langle \tilde n|(1-e^{i\omega_+}\hat U)^{-1}|0\rangle
    \langle 0|(1-e^{-i\omega_-}\hat U^\dagger)^{-1}|\tilde n\rangle\!
    \rangle_{\omega_0},
    \nonumber
\end{eqnarray}
where $\omega_\pm=\omega_0 \pm \frac{\omega}{2}$ and $\langle \cdot
\rangle_{\omega_0}\equiv \frac{1}{2\pi}\int_0^{2\pi}d\omega_0$.

\begin{figure}
\includegraphics[width=8.5cm]{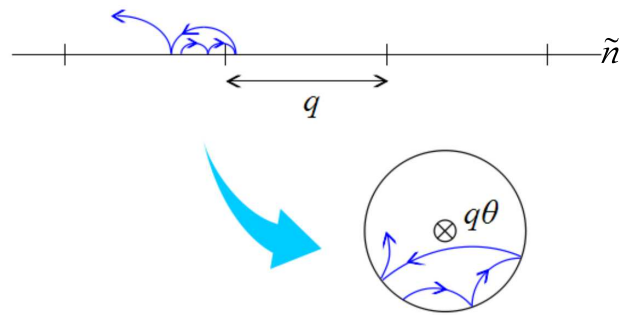}
\caption{(Color online) The translation symmetry entails a reduction of
quantum dynamics from unbounded angular momentum space to a circle of
circumference $q$ pierced by an Aharonov-Bohm like flux $q \theta$ with
$\theta$ being the Bloch angle.}
\label{fig:1}
\end{figure}

To proceed we consider the eigenvalues and eigenvectors of $\hat U$.
The translation symmetry (\ref{eq:12}) entails a good quantum number, i.e.,
the Bloch angle $\theta \in [0,2\pi/q]$. According to Bloch's theorem,
\begin{eqnarray}
\hat{U}|\psi_{\alpha,\theta} \rangle = e^{i \epsilon_\alpha(\theta)}|
\psi_{\alpha,\theta} \rangle,
\label{eq:3}
\end{eqnarray}
where both the quasienergy spectrum, $\{\epsilon_\alpha(\theta)\}$, and
corresponding eigenvectors, $\{|\psi_{\alpha,\theta} \rangle\}$, depend
on $\theta$. The Bloch wave function
\begin{equation}\label{eq:23}
\psi_{\alpha,\theta}(\tilde n)\equiv \langle \tilde n|\psi_{\alpha,\theta}
\rangle=e^{i\tilde n\theta}\varphi_{\alpha,\theta}(\tilde n),
\end{equation}
with $\varphi_{\alpha,\theta}(\tilde n)$ being a $q$-periodic wave function,
\begin{equation}\label{eq:20}
\varphi_{\alpha,\theta}(\tilde n)=\varphi_{\alpha,\theta}(\tilde n+q)=
\varphi_{\alpha,\theta}(n).
\end{equation}
In deriving the last equality of Eq.~(\ref{eq:20}) we use the fact that
$\tilde n$ can be uniquely written as $n+Nq$, where $0\leq n \leq q-1$ and
$N\in \mathbb{Z}$. Upon applying Bloch's theorem to Eq.~(\ref{eq:21})
we express the energy profile as
\begin{eqnarray}
E(t)=-\frac{1}{2}\int\frac{d\omega}{2\pi}e^{-i\omega t}{\sum_{n=0}^{q-1}}
\int_b d\theta_+ \partial_{\theta_-}^2|_{\theta_-=\theta_+}K_\omega(n),
\label{eq:16}
\end{eqnarray}
where the notation: $\int_b d\theta \equiv \frac{q}{2\pi}
\int_0^{2\pi/q} d\theta$ has been introduced and
\begin{eqnarray}\label{eq:22}
    K_\omega(n)&=&\langle\! \langle n|\frac{1}{1-e^{i\omega_+}
    \hat U_{\theta_+}}|0\rangle \nonumber\\
    &\times& \langle 0|\frac{1}{1-e^{-i\omega_-}\hat U^\dagger_{\theta_-}}
    |n\rangle\!\rangle_{\omega_0}
\end{eqnarray}
is the density correlation function for autonomous stroboscopic dynamics
restricted on a circle of circumference $q$. This reduced system is governed
by a $\theta$-dependent Floquet operator,
\begin{eqnarray}
\hat{U}_\theta = e^{-\frac{i}{2\tilde h}H_0}e^{-\frac{iK}{\tilde h}
\cos(\hat\theta+\theta)}e^{-\frac{i}{2\tilde h}H_0}.
\label{eq:5}
\end{eqnarray}
Physically, $K_\omega(n)$ probes interference between retarded and advanced
amplitudes corresponding to the reduced dynamics, and the Bloch angle $\theta$
introduces an ``Aharonov-Bohm flux'', $q\theta$, piercing through the circle
(see Fig.~\ref{fig:1}). Note that the unitarity of $\hat{U}_\theta$ now implies
$\sum_{n''=0}^{q-1}(\hat{U}_\theta)_{nn''}(\hat{U}_\theta^\dagger)_{n''n'}=
\delta_{nn'}$ for arbitrary $n,n'\in \{0,1,\cdots, q-1\}$.

It is important to note that for the reduced system governed by
$\hat{U}_\theta$ the effective time-reversal symmetry $\hat T_c$ is broken
by the Bloch angle if $\theta\neq 0,\,\pi$. A question naturally arises: Does
this broken $\hat T_c$ symmetry bring the system to the unitary class in the
RMT? As we will show in the following, the answer crucially depends on whether
$H_0(\hat n)$ exhibits the inversion symmetry $\hat T_i$. If the latter
symmetry is broken, then the system belongs to the unitary class. If not,
then $\hat{U}_\theta$ is invariant under the combined operation $\hat T_i
\hat T_c$ and this brings the reduced system back to the orthogonal class.

%$\hat T_i\hat \theta^{\rm T}\hat T_i=\hat \theta$, it is easy to show
%$[\cos(\hat\theta+\theta),\hat T_i\hat T_c]=0$,

%=====================Fig.1============================================================================
%=======================================================================================================

\subsection{Universal metal-supermetal dynamics crossover}
\label{sec:energy_growth}

\subsubsection{Field theory}
\label{sec:field_theory}

The remaining task is to calculate the two-particle Green's function
${K_\omega(n)}$ evaluated at fixed parameters $\theta_\pm$. Following
Ref.~\cite{Tian} we may express it in terms of a functional integral,
\begin{eqnarray}
K_\omega(n)&=&-\frac{1}{2^4} \int dQ\, e^{-S[Q]}\nonumber\\
&\times& {\rm str}\left(k
%(1+\Lambda)(1-\tau_3)
(Q(n))_{+2,-2}k(Q(0))_{-2,+2}\right).
\label{eq:17}
\end{eqnarray}
Here $Q\equiv \{Q_{\lambda \alpha \beta,\lambda' \alpha' \beta'}\}$ is
a supermatrix defined on three sectors: the index $\lambda=+,-$ refers
to the advanced and retarded (`AR') sector, $\alpha={f,\, b}$ to the
fermionic-bosonic (`FB') sector, and $\beta=1,2$ to the `T' sector
accommodating the symmetry under the combined operation $\hat T_i\hat T_c$;
i.e., $\beta=1$ ($2$) refers to (in)variance under the $\hat T_i\hat T_c$
operation. $Q$ satisfies the periodic boundary condition $Q(n)=Q(n+q)$.
The constant matrix $k=\sigma_{\rm FB}^3\otimes \sigma_{\rm AR}^0 \otimes
\sigma_{\rm T}^0$, where $\sigma_{\rm X}^0$ is the unit ($2\times 2$)
matrix in the X sector. The action is given by
\begin{eqnarray}
\label{eq:18}
  S = \frac{1}{2^4}\int_0^q dn {\rm str}\big(D_q(i\partial_n
  Q-[\hat \vartheta,Q])^2-2i\omega Q\sigma_{\rm AR}^3\big)
\end{eqnarray}
with
\begin{equation}\label{eq:19}
    \hat \vartheta=\left(
         \begin{array}{cc}
         \theta_+ & 0 \\
         0 & \theta_- \\
         \end{array}
   \right)_{\rm AR}\otimes \sigma_{\rm T}^{-1+\beta^2}\otimes
   \sigma_{\rm FB}^0.
\end{equation}
For $\beta=2$ the matrix $\sigma_{\rm T}^{-1+\beta^2}=\sigma_{\rm T}^3$
represents the sign change in the Aharonov-Bohm flux under the $\hat T_c$
operation. The corresponding effective field theory has been obtained
in Ref.~\cite{Tian}, but for $\beta=1$, an additional symmetry,
\begin{equation}\label{eq:33}
    Q(n)=Q(q-n),
\end{equation}
arises.
Note that the above action is universal in the sense that
it depends on system parameters only through the diffusion constant
$D_q$. The effective field theory is valid, provided the parametric
conditions, (i) $1\ll K/\tilde h \ll q \ll (K/\tilde h)^2$, (ii) $\omega
\ll 1$, and (iii) $K\gg 1$ \cite{note5}, are met. The physical meanings
of these conditions are as follows. For (i), $K/\tilde h$ plays the role
of the ``transport mean free path'' in normal metals; the inequality
$K/\tilde h \ll q$ guarantees the validity of the hydrodynamic expansion,
while $q \ll (K/\tilde h)^2$ implies that localization physics does not
play any roles. For (ii), means that we are concerned in time scales much larger
than the kicking period. In (iii) and $K/\tilde h\gg 1$, it is implied that the angular
(or more precisely, $\sin\hat\theta$) correlation rapidly decreases, i.e.,
strong chaoticity. The latter results in the fact that the distribution of the
quasienergy spectrum $\{\epsilon_\alpha(\theta)\}$ of the Floquet operator
$\hat{U}_\theta$ follows the Wigner-Dyson statistics for the circular
orthogonal ensemble (COE) for $\beta=1$ and the circular unitary ensemble
(CUE) for $\beta=2$.

In principle, the diffusion constant $D_q$ depends on both the parameters
$K$ and $\tilde h$ and the details of $H_0$, i.e., $c_k$. The $H_0$ dependence
arises from the short-time correlation contributions to the diffusion
constant~\cite{Tian}. For $K/\tilde h \gg 1$ these correlation contributions
are negligible and consequently only the parameter ($K,\tilde h$) dependence
arises, i.e., $D_q = D_q(K, \tilde h)\sim(K/\tilde h)^2$.

%=====================Fig.2============================================================================
\begin{figure}
\includegraphics[width=8.7cm] {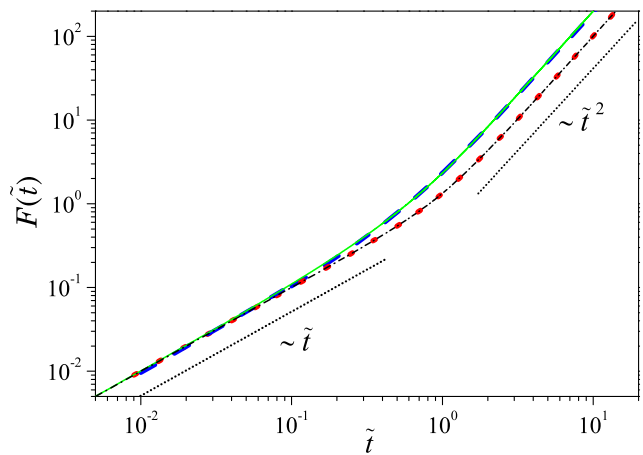}
\caption{(Color online) The analytical results obtained by the field
theory for a generalized kicked rotor with (thick blue dashed line) and
without (thick red dotted line) the inversion symmetry. It shows that the
energy profile exhibits a universal crossover from metallic ($\sim \tilde
t$) to supermetallic ($\sim \tilde t^2$) growth. The results are in excellent
agreement with those obtained from the RMT corresponding to the orthogonal
(thin green solid line) and the unitary (thin black dash-dotted line)
symmetry, respectively.}
\label{fig:2}
\end{figure}
%=======================================================================================================

\subsubsection{Metallic-supermetallic growth crossover}
\label{sec:explicit_results}

We proceed to use Eqs.~(\ref{eq:16}), (\ref{eq:17}), and (\ref{eq:18}) to
explicitly calculate $E(t)$. First of all, for $\omega\lesssim \Delta\equiv
2\pi/q$ which is the mean level spacing of the unit cell, the effective action
(\ref{eq:18}) shows that inhomogeneous $Q$-field fluctuations have a typical
action $E_c/\Delta \gg1$ where the Thouless energy $E_c = D_q/q^2$ is the
inverse of the classical diffusion time through the periodicity volume. This
inequality is ensured by $q\ll (K/\tilde h)^2\sim D_q$. So, inhomogeneous
fluctuations can be neglected and only the zero mode, $Q(n)\equiv Q$, is kept.
Equation (\ref{eq:18}) thereby is simplified to the zero mode action,
\begin{eqnarray}
S=\frac{\pi}{8\Delta}{\rm str}(D_q[\hat{\vartheta},Q]^2-2i\omega Q\sigma_{\rm AR}^3).
\label{eq:27}
\end{eqnarray}
Then, we substitute this action into Eq.~(\ref{eq:17}). The subsequent
calculations are similar to those of Ref.~\cite{Tian} and below we will give
the results. Some details are given in Appendix~\ref{sec:derivation}. We find
\begin{equation}
\label{eq:26}
    E(t)/(qD_q)=F(\tilde{t}),\quad \tilde{t}=t/q.
\end{equation}
It is important that this energy profile is universal: the details of the model
[e.g., the stochastic parameter $K$, the free Hamiltonian $H_0$, the denominator
of rational $\tilde h/(4\pi)$, etc.] only determine the scales of energy and
time. The explicit form of the universal function $F(\tilde t)$ depends on
the system's symmetry and is given as follows:
%\begin{itemize}
%\item

%=====================Fig.3============================================================================
\begin{figure}
\vskip-0.038cm
\includegraphics[width=8.7cm]{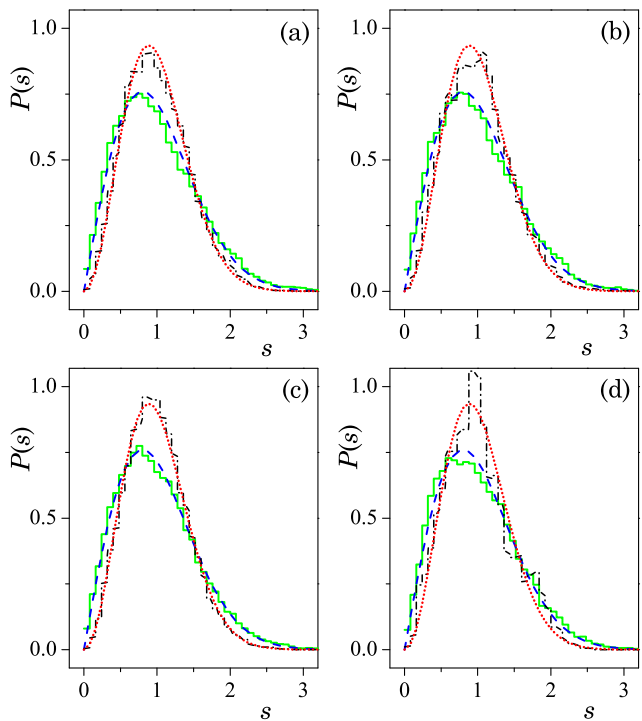}
\caption{(Color online) The simulation results of the distribution of the nearest
level spacing $P(s)$ for $\alpha_2=1,\alpha_3=0$ (green solid histograms)
and $\alpha_2=0, \alpha_3=1$ (black dash-dotted histograms) follow the Wigner-Dyson statistics for COE (blue dashed lines) and CUE
(red dotted lines), respectively. The parameter $\tilde h/(4\pi)$ is
$55/691$ (a), $24/301$ (b), $24/299$ (c), and $7/87$ (d). For all four panels
$K=300$.}
\label{fig:3}
\end{figure}
%=======================================================================================================

$\bullet$~For $H_0$ with inversion symmetry so that the quantum system reduced
to a unit cell is orthogonal, we find (see Appendix~\ref{sec:derivation} for
derivations)
\begin{eqnarray}\label{eq:7}
F\left(\tilde{t}\right)&=&\frac{1}{8}\int_1^\infty d\lambda_1\int_1^\infty d
\lambda_2\int_{-1}^1 d\lambda \delta(2\tilde{t}+\lambda-\lambda_1\lambda_2)
\nonumber\\
&\times& \frac{(1-\lambda^2)(1-\lambda^2-\lambda_1^2-\lambda_2^2+2\lambda_1^2
\lambda_2^2)^2}{(\lambda^2 + \lambda_1^2 + \lambda_2^2 - 2\lambda\lambda_1
\lambda_2-1)^2}.
\end{eqnarray}
For short times ($\tilde t\ll 1$) the Dirac function in Eq.~(\ref{eq:7})
implies that the integrals are dominated by $\lambda, \lambda_{1,2}\approx 1$.
Therefore, we may expand $\lambda,\lambda_{1,2}$ near unity and keep the leading
expansion of $F(\tilde t)$. As a result, $F(\tilde t\ll 1)\sim \tilde t$. For
long times ($\tilde t\gg 1$), the integrals are dominated by $\lambda={\cal O}
(1)$ and $\lambda_{1,2}\gg 1$. Taking this into account we find $F(\tilde t
\gg 1) \sim \tilde t^2$.
%\item

$\bullet$~For $H_0$ without the inversion symmetry so that the quantum system
reduced to a unit cell is unitary, the result is the same as what has been
found previously~\cite{Tian}, which is
\begin{eqnarray}\label{eq:8}
F\left(\tilde{t}\right)=\bigg\{
\begin{array}{rcl}
\tilde{t}+\frac{1}{3}\tilde{t}^3,&0<\tilde{t}<1,\\
\tilde{t}^2+\frac{1}{3},&\tilde{t}>1.
\end{array}
\end{eqnarray}
%\end{itemize}
%\noindent

Eqs.~(\ref{eq:7}) and (\ref{eq:8}) show that the energy
profile exhibits a universal crossover from metallic to supermetallic
growth (Fig.~\ref{fig:2}). As we will show in Sec.~\ref{sec:discussion},
they can be obtained also from RMT.

\section{\label{sec:simuresults} Numerical tests}

In this section we put the analytic results, 
namely, Eqs.~(\ref{eq:7}) and (\ref{eq:8}), to numerical tests. As we
will see below, although the analytical derivations of these two universal
crossovers require the conditions $1\ll K/\tilde h\ll q \ll (K/\tilde h)^2$
and $K\gg 1$, they hold in broader regimes. Surprisingly, they hold even in
systems (e.g., the kicked Harper system~\cite{KH1, KH2, KH3}) the quantum
dynamical behaviors of which at irrational $\tilde h/(4\pi)$ are fundamentally
different from the standard kicked rotor~\cite{KH4, KH5}.

To numerically study the universality of the dynamics crossover we
use different forms of the free Hamiltonian $H_0$. It has been known that
this Hamiltonian, $H_0$, mimics ``impurities'' in genuine disordered
systems~\cite{Fishman, Tian}.

\subsection{\label{sec:polynomial} Polynomial-type free Hamiltonian}

We first consider a polynomial-type free Hamiltonian with a specific form
given by
\begin{eqnarray}\label{eq:1}
H_0 = \frac{\tilde h^2}{2}\left(\alpha_2 \hat{n}^{2}+\alpha_3 \hat{n}^{3}\right),
\end{eqnarray}
with $\alpha_{2,3} \in \mathbb{N}$. The ${\hat T}_i$ symmetry is present
for vanishing $\alpha_3$ and otherwise broken.

\subsubsection{Spectral statistics}
\label{sec:spectral_statistics}

%=====================Fig.4============================================================================
\begin{figure}[!]
\includegraphics[width=8.7cm]{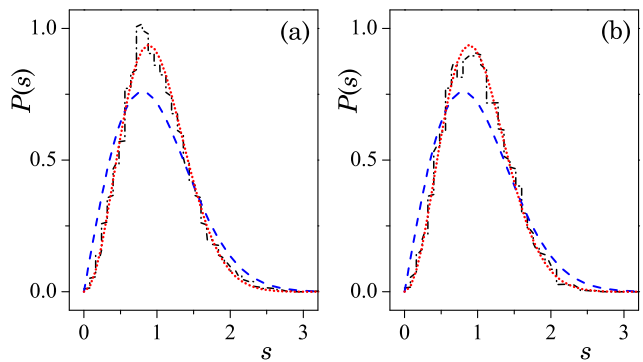}
\caption{(Color online) The simulation results of $P(s)$ (black dash-dotted
histograms) follow the Wigner-Dyson statistics for CUE (red dotted line), as long as $\alpha_3$ does not vanish. For
panel (a) $\alpha_2=1$ and $\alpha_3=99$ while for (b) $\alpha_2=99$ and
$\alpha_3=1$. In both cases $\tilde h/(4\pi)=8/101$ and $K=300$. The blue
dashed lines are for the Wigner-Dyson statistics for COE.}
\label{fig:4}
\end{figure}
%=======================================================================================================

For a given Bloch angle $\theta$ we numerically diagonalize the Floquet operator
$\hat{U}_\theta$ to obtain the quasienergy spectrum $\{\epsilon_\alpha(\theta)\}$.
We repeat the computation for $1000$ randomly selected values of $\theta$ so that
a large ensemble is realized. We calculate the distribution of the nearest
level spacing, $P(s)$, for various parameters of $\alpha_{2,3}, K$ and $\tilde h$.
We find that $P(s)$ follows the Wigner-Dyson statistics
for COE for vanishing $\alpha_3$ (Fig.~\ref{fig:3}, green solid histograms) and CUE for
vanishing $\alpha_2$ (Fig.~\ref{fig:3}, black dash-dotted histograms). More generally, as
long as $\alpha_3\neq 0$, $P(s)$ follows the Wigner-Dyson statistics for CUE (Fig.~\ref{fig:4}).
These results indicate that the system is strongly chaotic so that the field
theory is expected to be valid.

%=====================Fig.5============================================================================
\begin{figure}
\vskip-0.038cm
\includegraphics[width=8.7cm]{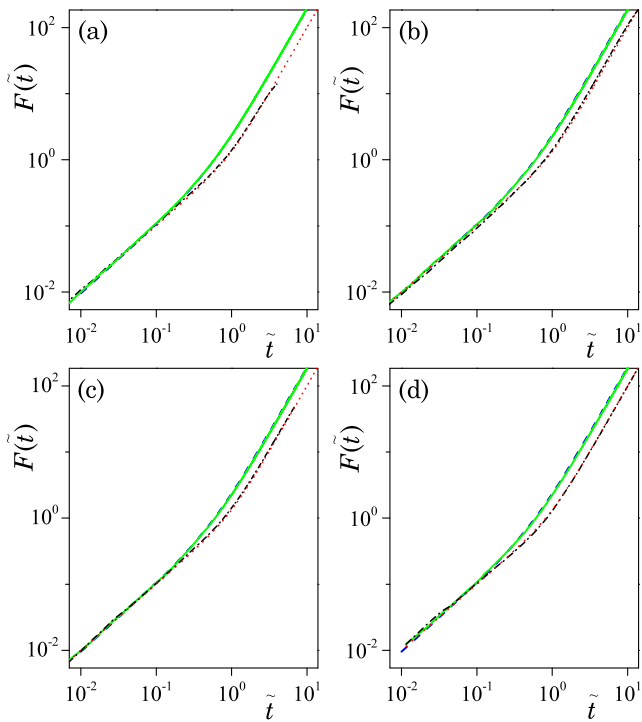}
\caption{(Color online) The simulation results of the rescaled energy
growth $F(\tilde t)$ for $\alpha_2=1,\alpha_3=0$ (green solid lines) and
$\alpha_2=0, \alpha_3=1$ (black dash-dotted lines) convincingly support
the analytic predictions for orthogonal (blue dashed lines) and unitary
(red dotted lines) systems, respectively. The parameters for each panel
are the same as in Fig.~\ref{fig:3}.}
\label{fig:5}
\end{figure}
%=======================================================================================================

\subsubsection{Universal energy growth}
\label{sec:diffusive-ballistic_crossover}

\begin{figure}[!]
\includegraphics[width=8.7cm] {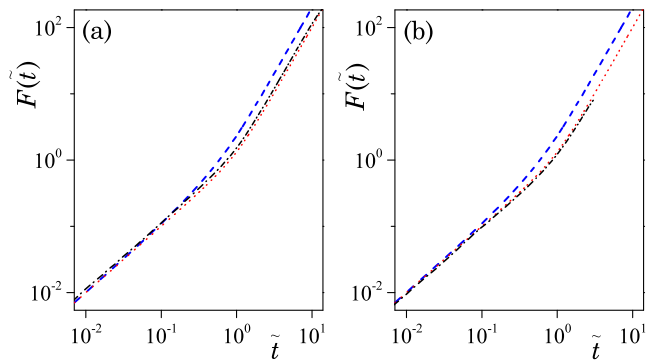}
\caption{(Color online) The simulation results (black dash-dotted lines)
show that as long as $\alpha_3$ does not vanish, $F(\tilde t)$ follows
the metallic-supermetallic growth crossover of unitary type described
by  Eq.~(\ref{eq:8}) (red dotted lines). The parameters for each panel
are the same as in Fig.~\ref{fig:4}. For comparison, the crossover of
orthogonal type described by Eq.~(\ref{eq:7}) (blue dashed lines) is
also plotted.}
\label{fig:6}
\end{figure}

For a given free Hamiltonian $H_0$, we use the standard fast Fourier
transform technique to simulate the quantum evolution (\ref{eq:6}) to
compute $E(t)$ defined by Eq.~(\ref{Etdf}) and in turn $F(\tilde{t})$
by Eq.~(\ref{eq:26}).
Fig.~\ref{fig:5} presents the simulation results of $F(\tilde t)$ for
the free Hamiltonian $H_0$ given by Eq.~(\ref{eq:1}) with the same parameters
as in Fig.~\ref{fig:3} for the sake of comparison. We find that for vanishing
$\alpha_3$ where the ${\hat T}_i$ symmetry is present, $F(\tilde{t})$ is
in excellent agreement with the analytic result given by Eq.~(\ref{eq:7}),
while for vanishing $\alpha_2$ where the ${\hat T}_i$ symmetry is broken,
$F(\tilde{t})$ is in excellent agreement with the analytic result given by
Eq.~(\ref{eq:8}). We also find that the universal metallic-supermentallic
growth crossovers (\ref{eq:7}) and (\ref{eq:8}) are valid not only in the
regime of $K/\tilde h\ll q \ll (K/\tilde h)^2$ (Fig.~\ref{fig:5}(a)), but
also of $q \lesssim K/\tilde h$ (Fig.~\ref{fig:5}(b)-(d)). In
addition, as shown in Fig.~\ref{fig:6}, as long as $\alpha_3\neq 0$, for
which the quasienergy spectrum follows the Wigner-Dyson statistics for
CUE (see Fig.~\ref{fig:4}), the energy growth follows the
metallic-supermetallic growth crossover of unitary type described by
Eq.~(\ref{eq:8}), confirming the sensitivity of the crossover to
the system's symmetry.

%=====================Fig.6============================================================================
%=======================================================================================================

\subsection{\label{sec:KHM} Nonpolynomial-type free Hamiltonian}

When the polynomial is of higher degree (with all coefficients being
integers) but finite, we find numerically that the system's behavior at
resonance remains the same. A natural question thereby arises: What happens
if the degree is infinite so that $H_0$ is not a polynomial? To investigate
this problem we consider
\begin{eqnarray}\label{eq:9}
H_0=L\cos(\tilde h\hat{n}+\phi).
\end{eqnarray}
Here $L$ is a (nonzero) constant. When the phase parameter $\phi$ vanishes,
this system becomes the conventional kicked Harper model~\cite{KH1,KH2,KH3}
and exhibits the inversion symmetry. Whereas $\phi\neq 0$ the inversion
symmetry breaks.

\begin{figure}
\includegraphics[width=8.7cm] {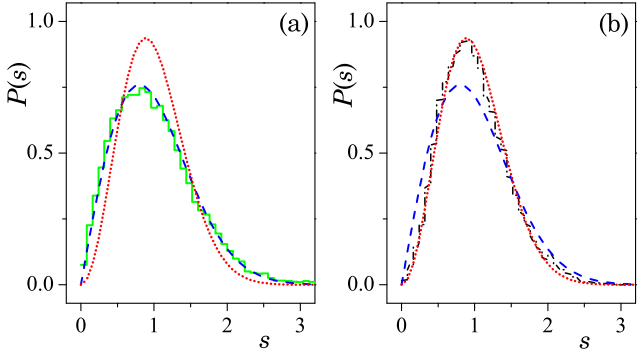}
\caption{(Color online) The simulation results of $P(s)$ for the
kicked Harper model. The green solid histogram for $\phi=0$ in (a)
and the black dash-dotted histogram for $\phi=\pi/2$ in (b) follow
the Wigner-Dyson statistics for COE (blue dashed line) and
CUE (red dotted line), respectively. For both panels
$\tilde h/(4\pi)=24/301$ and $L=K=300$.}
\label{fig:7}
\end{figure}

\begin{figure}
\includegraphics[width=8.7cm] {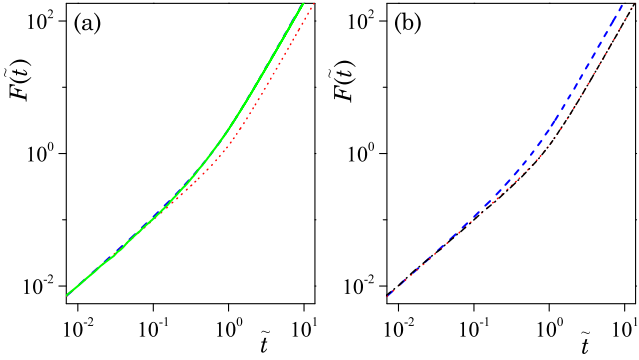}
\caption{(Color online) The simulation results, green solid line
in (a) and black dash-dotted line in (b), of the energy growth
corresponding to the two cases in Fig.~\ref{fig:7}. They exhibit
again the universal metallic-supermetallic growth crossover as
analytically predicted by Eq.~(\ref{eq:7}) for the orthogonal
(blue dashed line) and by Eq.~(\ref{eq:8}) for the unitary (red
dotted line) symmetry.}
\label{fig:8}
\end{figure}

As shown in Fig.~\ref{fig:7}, the spectral fluctuations of the Floquet
operator $\hat U_\theta$ obey the Wigner-Dyson statistics for COE (CUE) for $\phi=0$ ($\pi/2$). Correspondingly, $F(\tilde{t})$ is
in excellent agreement with the analytic result given by Eq.~(\ref{eq:7})
for $\phi=0$ (Fig.~\ref{fig:8}(a)) and by Eq.~(\ref{eq:8})
(Fig.~\ref{fig:8}(b)) for $\phi=\pi/2$ . Therefore, the analytic results
for universal metal-supermetal dynamics crossover are also confirmed for
the kicked Harper model.

%=====================Fig.7============================================================================
%=======================================================================================================

\section{\label{sec:discussion}Dynamics universality from RMT}

In this section we provide an alternative scheme for approximate analytic
calculations of the energy profile. From this scheme we will see that
the universality of the dynamics crossover is deeply rooted in the universal
fluctuations of the Bloch's bands and eigenfunctions which are well
described by the RMT. Interestingly, this scheme unveils a connection
of the quantum resonance (supermetallic phase) of kicked rotors to the
periodic multi-baker map~\cite{Dorfman1, Dorfman2, Dorfman}.

\subsection{General expression of $E(t)$}
\label{sec:energy_profile}

To this end we compactify the angular momentum space with the periodic
boundary condition so that it includes $M$ unit cells. Recall that the
initial condition is not essential. We consider the energy growth with
the initial state $|\psi(0)\rangle=|\tilde n\rangle$ and average the
energy profile with respect to $\tilde n$.
%(For simplicity we make the dynamics effectively start from the kicking.
%To this end, we introduce the phase factor $e^{\frac{i}{2\tilde h}
%H_0(\tilde n)}$ so that the last factor in Eq.~(\ref{eq:4}) is compensated.)
Mathematically, this is equivalent to preparing an ``equilibrium state'',
$\rho_{eq}=(qM)^{-1}\sum_{\tilde n} |\tilde n\rangle\langle \tilde n|$.
%obeying $\hat U \rho_{eq}\hat U^\dagger =\rho_{eq}$.

%=====================Fig.8============================================================================
%=======================================================================================================

By the definition of $E(t)$ we find
\begin{eqnarray}\label{eq:34}
    E(t) = \frac{1}{2}\sum_{\tau_{1,2}=0}^{t-1}{\rm Tr}(\rho_{eq}
    \hat v(\tau_1) \hat v(\tau_2))
    \equiv \frac{1}{2}\sum_{\tau_{1,2}=0}^{t-1} C_{\tau_1,\tau_2},
\end{eqnarray}
where $\hat v(\tau)\equiv \frac{K}{\tilde h}\sin\hat\theta(\tau)$ may
be considered as the velocity (corresponding to the motion in angular
momentum space) operator, with
\begin{equation}\label{eq:35}
    \hat v(\tau) = (\hat U^\dagger)^\tau \hat v\, \hat U^\tau,
\end{equation}
and $C_{\tau_1,\tau_2}$ is its auto-correlation function. Equation (\ref{eq:34})
is further reduced to
\begin{eqnarray}\label{eq:36}
    E(t)= \frac{1}{2}\left(C_0 t+2\sum_{\tau=1}^{t-1} (t-\tau)C_{\tau}\right)
\end{eqnarray}
thanks to $C_{\tau_1,\tau_2}=C_{\tau_1-\tau_2}$.

Next, we calculate the auto-correlation function $C_\tau$. It can be
rewritten as
\begin{eqnarray}\label{eq:37}
    C_\tau=\left(\frac{K}{\tilde h}\right)^{2}\frac{1}{qM}{\rm Tr}
    ((\hat U^\dagger)^\tau \sin\hat \theta\hat U^\tau \sin\hat \theta)
\end{eqnarray}
by using its definition [cf. Eq.~(\ref{eq:34})]. Employing the Bloch's
theorem (\ref{eq:3}) we obtain
\begin{eqnarray}
    C_\tau &=& \left(\frac{K}{\tilde h}\right)^{2}\frac{1}{qM}\sum_{\alpha,
    \alpha'}\sum_{\theta,\theta'} e^{-i(\epsilon_\alpha(\theta)-
    \epsilon_{\alpha'}(\theta'))\tau}\nonumber\\
    &\times&|\tilde {\bf J}_{\alpha\theta,\alpha'\theta'}|^2,\label{eq:38}\\
    \tilde {\bf J}_{\alpha\theta,\alpha'\theta'} &=& \langle \psi_{\alpha,\theta}
    |\sin\hat\theta|\psi_{\alpha',\theta'}\rangle.
    \label{eq:39}
\end{eqnarray}
The matrix elements of $\tilde {\bf J}$ can be simplified to
\begin{eqnarray}
\tilde {\bf J}_{\alpha\theta,\alpha'\theta'}&=&\frac{1}{2iM}\sum_{\tilde n,\tilde n'}
\varphi_{\alpha,\theta}^*(\tilde n)\varphi_{\alpha',\theta'}(\tilde n')e^{i(\tilde n
\theta-\tilde n'\theta')}\nonumber\\
&&\times\left(\delta_{\tilde n-\tilde n',1}-\delta_{\tilde n-\tilde n',-1}\right)\nonumber\\
&=&\frac{1}{2i}\delta_{\theta\theta'}\sum_{n=0}^{q-1}(\varphi_{\alpha,\theta}^*(n)
\varphi_{\alpha',\theta}(n-1)e^{i\theta}\nonumber\\
&&-
\varphi_{\alpha,\theta}^*(n)\varphi_{\alpha',\theta}(n+1)e^{-i\theta})\nonumber\\
&\equiv& \delta_{\theta\theta'}\tilde {\bf J}_{\alpha\alpha'}(\theta).
\label{eq:40}
\end{eqnarray}
In deriving the second equality we have summed over the Bravis lattice which
enforces $\theta=\theta'$. Substituting Eq.~(\ref{eq:40}) into Eq.~(\ref{eq:38})
gives
\begin{eqnarray}
    C_\tau = \left(\frac{K}{\tilde h}\right)^{2}\frac{1}{qM}\sum_{\alpha,\alpha'}
    \sum_{\theta} e^{-i(\epsilon_\alpha(\theta)-\epsilon_{\alpha'}(\theta))\tau}
    |\tilde {\bf J}_{\alpha\alpha'}(\theta)|^2.
    \label{eq:41}
\end{eqnarray}
By setting $\tau=0$ in Eqs.~(\ref{eq:37}) and (\ref{eq:41}) we find the
following identity,
\begin{eqnarray}
    \frac{1}{Lq}\sum_{\alpha,\alpha'}\sum_{\theta} |\tilde
    {\bf J}_{\alpha\alpha'}(\theta)|^2=\frac{1}{2}.
    \label{eq:46}
\end{eqnarray}
Eqs.~(\ref{eq:36}) and (\ref{eq:41}) constitute an exact formalism
for calculating $E(t)$. Formally, they are identical to the formulas giving
the mean-squared displacement in the periodic multi-baker map.
The latter essentially is composed of an infinite number of standard baker's maps,
which are coupled to each other, with the coupling being spatially periodic~\cite{note3}.
(The details are certainly not.)

\subsection{Effects of chaotic fluctuations of Bloch bands and wave functions}
\label{sec:chaotic_fluctuation}

Equation (\ref{eq:41}) shows that the correlation is governed by the
quasienergy spectrum $\{\epsilon_\alpha(\theta)\}$ and the corresponding
eigenfunctions $\varphi_{\alpha,\theta}$, which determine the oscillatory
factor $e^{-i(\epsilon_\alpha(\theta)-\epsilon_{\alpha'}(\theta))\tau}$
and the matrix elements of the velocity operator, i.e., $\tilde
{\bf J}_{\alpha\alpha'}(\theta)$, respectively. Because the reduced
quantum system in a unit cell (for given Bloch angle $\theta$) is
chaotic, both $\{\epsilon_\alpha(\theta)\}$ and $\varphi_{\alpha,\theta}$
exhibit chaotic fluctuations. If we assume that the fluctuations of
$\{\epsilon_\alpha(\theta)\}$ and $\varphi_{\alpha,\theta}$ are
independent, then the correlation function is simplified to
\begin{eqnarray}
    C_\tau &\rightarrow& \left(\frac{K}{\tilde h}\right)^{2}\frac{1}{Lq}
    \sum_{\alpha,\alpha'}\sum_{\theta} \left\langle e^{-i(\epsilon_\alpha
    (\theta)-\epsilon_{\alpha'}(\theta))\tau}\right\rangle \nonumber\\
    &&\times \left\langle |\tilde {\bf J}_{\alpha\alpha'}(\theta)|^2
    \right\rangle,
    \label{eq:42}
\end{eqnarray}
where $\langle\cdot\rangle$ stands for the averages over a random ensemble
of energy spectrum and wave functions, respectively. These random ensembles
are well described by the RMT.

For the system without the inversion symmetry ($\beta=2$), following
Ref.~\cite{Dorfman} we use the RMT corresponding to CUE to find
\begin{eqnarray}
\left\langle e^{-i(\epsilon_\alpha(\theta)-\epsilon_{\alpha'}(\theta))
\tau}\right\rangle=\Bigg\{\begin{array}{c}
         1,\quad \tau=0; \\
         \frac{\tau-q}{q(q-1)},\quad 0<\tau<q; \\
         0. \quad \tau\geq q.
       \end{array}
\label{eq:43}
\end{eqnarray}
For the system with the inversion symmetry ($\beta=1$), we find
\begin{eqnarray}
&&\left\langle e^{-i(\epsilon_\alpha(\theta)-\epsilon_{\alpha'}(\theta))
\tau}\right\rangle\nonumber\\
&=&\Bigg\{\begin{array}{c}
         1,\, \tau=0; \\
         \frac{1}{q(q-1)}\left(-q+2\tau \left(f(\frac{q}{2}+\tau)-
         f(\frac{q}{2})\right)\right),\, 0<\tau<q; \\
         \frac{1}{q(q-1)}\left(-q+2\tau \left(f(\frac{q}{2}+\tau)-
         f(\tau-\frac{q}{2})\right)\right),\, \tau\geq q
       \end{array}
       \nonumber\\
\label{eq:44}
\end{eqnarray}
corresponding to COE, where $f(\tau)\equiv
\sum_{k=1}^\tau \frac{1}{2k-1}$. Note that these results are independent
of $\theta$.
%In addition, their dependence of the RMT universality class has confirmed
%by simulations as shown above.

We cannot calculate the second factor in Eq.~(\ref{eq:42}) analytically.
Instead, motivated by the similarity between Eq.~(\ref{eq:41}) and the
corresponding expression for the periodic multi-baker map, we hypothesize
that the matrix elements of the velocity operator have the universal
behavior. So, we translate the results of Ref.~\cite{Dorfman} obtained
from the RMT to the present context which read
\begin{eqnarray}
\left\langle |\tilde {\bf J}_{\alpha\alpha'}(\theta)|^2\right\rangle=
\bigg\{\begin{array}{c}
         \frac{1}{2}\frac{3-\beta}{q+3-\beta},\quad \alpha=\alpha'; \\
         \frac{1}{2}\frac{q}{(q-1)(q+3-\beta)},\quad \alpha\neq \alpha',
         \end{array}
\label{eq:45}
\end{eqnarray}
where the overall factor of $1/2$ makes it obey the identity (\ref{eq:46}).
Below key predictions obtained from this ansatz are confirmed.

We substitute Eqs.~(\ref{eq:43})-(\ref{eq:45}) into Eq.~(\ref{eq:42}).
In combination with Eq.~(\ref{eq:36}) we obtain for $\beta=1$
($D_q=(\frac{K}{2\tilde h})^2$)
\begin{eqnarray}
\frac{E(t)}{qD_q}=\bigg\{\begin{array}{c}
  \tilde t + \frac{\tilde t (\tilde t-q^{-1})}{(1+2q^{-1})}\left(1+
  \frac{\tilde t-2q^{-1}}{3(1-q^{-1})}\right),\quad 0<\tilde t<1; \\
  \frac{2\tilde t^2}{1+2q^{-1}}+\frac{1+q^{-1}}{3(1+2q^{-2})},
  \quad \tilde t\geq 1,
  \end{array}
\label{eq:48}
\end{eqnarray}
where small corrections have been ignored, and for $\beta=2$
\begin{eqnarray}
\frac{E(t)}{qD_q}=\bigg\{\begin{array}{c}
  \tilde t + \frac{\tilde t (\tilde t-q^{-1})(\tilde t-2q^{-1})}
  {3(1-q^{-2})},\quad 0<\tilde t<1; \\
  \frac{\tilde t^2}{1+q^{-1}}+\frac{1}{3}, \quad \tilde t\geq 1.
  \end{array}
\label{eq:47}
\end{eqnarray}
As shown in Fig.~\ref{fig:2}, these results are in excellent agreement
with those predicted by the field theory and are confirmed by simulations
as shown in Sec.~\ref{sec:simuresults}. Indeed, for $\beta=2$,
Eq.~(\ref{eq:47}) is identical to the analytic expression (\ref{eq:8})
for $q\gg 1$. Equations (\ref{eq:48}) and (\ref{eq:47}) show that
\begin{equation}\label{eq:49}
\frac{E|_{\beta=1}(t)}{E|_{\beta=2}(t)}\stackrel{t\gg 1}
{\longrightarrow} 2,\quad {\rm for}\, q \gg 1.
\end{equation}

Summarizing, the universal metal-supermetal dynamics crossover can
be attributed to universal chaotic fluctuations of Bloch bands and
eigenfunctions. Since the latter are well described by the RMT, we
expect that such dynamics universality is identical to the RMT
universality. On the other hand, it is well known~\cite{Altland97}
that very rich RMT universality classes are brought about by different
symmetries. Although in this work only crossovers corresponding to the
orthogonal and unitary classes are found, the richness of RMT universality
classes implies that even richer universal crossover behaviors may exist,
provided the system exhibits other symmetries.

\section{Conclusion}
\label{sec:conclusion}

We analytically and numerically studied the dynamics of a large
class of generalized kicked rotor systems the Hamiltonian of which is given by
Eq.~(\ref{eq:13}) with rational $\tilde h/(4\pi)=p/q$. The universal laws
for the crossover from metal to supermetal dynamics are found, manifesting
themselves in universal crossovers from metallic (linear) to supermetallic
(quadratic) energy growth. These crossover behaviors are determined only by
the system's symmetry and insensitive to the details such as the values of
$p$ and $q$ and the strength of the kicking potential which only affect the
energy and the time scales. Specifically, we find that when the Hamiltonian
$H_0(\hat n)$ governing the free rotation has inversion symmetry, i.e.,
$H_0(\hat n)=H_0(-\hat n)$, the energy profile follows an orthogonal-type
crossover behavior, described by Eq.~(\ref{eq:7}), and otherwise the
crossover is a unitary type, described by Eq.~(\ref{eq:8}). We show by the
RMT that these universal dynamics crossovers can be attributed to the
universal Bloch wave functions and band fluctuations of quantum systems
reduced to a unit cell. This leads us to conjecture that in more general
chaotic systems with periodic driving, the universality class of the
metal-supermetal dynamics crossover is identical to that of eigenfunction
and spectral fluctuations described by the RMT.

Our results can be generalized to condensed-matter systems (such as
semiconductors) where Bloch bands have been seen to exhibit chaotic
fluctuations following the RMT. For these systems even richer universality
classes have been predicted~\cite{Altland97}. The ensuing metal-supermetal
dynamics crossovers are expected to fall into different universality classes
manifesting in different universal behaviors of the optical conductivity.
Provided the unit cell is so complicated that the Bloch energy spectrum
exhibits chaotic fluctuations following the Wigner-Dyson statistics, our
results are expected to be valid.

Finally, we remark that in Ref.~\cite{Guarneri09} the supermetallic growth
has been rigorously established for the standard kicked rotor under general
parametric conditions. This result was obtained without resorting to the
chaoticity condition. A byproduct of the present work, i.e., $E(t\rightarrow
\infty) \sim t^2$, is consistent with this rigorous result, but is established
based on the chaoticity condition. We believe that our main result of universal
metal-supermetal dynamics crossovers is of chaos origin. Such a crossover
phenomenon was not studied in Ref.~\cite{Guarneri09}.

\section*{Acknowledgements}
This work is supported by the National Natural Science Foundation of China (Grant Nos. 11174174, 11275159, 11335006,
and 11535011).

\begin{appendix}
\section{\label{sec:derivation}Derivations of Eq.~(\ref{eq:7})}

In this Appendix we derive Eq.~(\ref{eq:7}). To this end we employ the
polar coordinate representation of $Q$ \cite{Efetov97}. Specifically, we parametrize $Q$ as
\begin{eqnarray}
Q=RQ_0R^{-1},\quad Q_0=\left(
\begin{array}{rcl}
\cos\hat{\Theta} &i\sin\hat{\Theta}\\
-i\sin\hat{\Theta} &-\cos\hat{\Theta}
\end{array}
\right)_{\rm AR},
\label{eq:28}
\end{eqnarray}
where $Q_0$ is the radial part and
\begin{eqnarray}
\hat{\Theta}&=&\left(
\begin{array}{rcl}
\hat{\theta}_{11} &0\\
0 &\hat{\theta}_{22}
\end{array}
\right)_{\rm BF},\label{eq:29}\\
\hat{\theta}_{11}&=&\left(
\begin{array}{rcl}
\tilde{\theta} &0\\
0 &\tilde{\theta}
\end{array}
\right)_{\rm T},\quad
\hat{\theta}_{22}=i\left(
\begin{array}{rcl}
\tilde{\theta}_1 &\tilde{\theta}_2\\
\tilde{\theta}_2 &\tilde{\theta}_1
\end{array}
\right)_{\rm T}
\nonumber
\end{eqnarray}
with ${0<\tilde{\theta}<\pi,\tilde{\theta}_{1,2}>0}$, and $R$ is the
transverse part commuting with $\sigma_{\rm AR}^3$. It is important to
note that the Grassmann variables enter only into $R$ (see
Ref.~\cite{Efetov97} for details).

We substitute this parametrization into Eqs.~(\ref{eq:17}) and (\ref{eq:27}).
Taking into account that Eq.~(\ref{eq:19}) now is simplified to
\begin{equation}\label{eq:32}
    \hat \vartheta=\left(
                     \begin{array}{cc}
                       \theta_+ & 0 \\
                       0 & \theta_- \\
                     \end{array}
       \right)_{\rm AR}\otimes \sigma_{\rm T}^{0}\otimes \sigma_{\rm FB}^0,
\end{equation}
we find
\begin{eqnarray}
K_\omega&=&\frac{1}{4} \int_1^\infty d\lambda_1 \int_1^\infty d\lambda_2
\int_{-1}^1 d\lambda \, e^{-S} \nonumber\\
&\times&\frac{(1-\lambda^2)(1-\lambda^2-\lambda_1^2-\lambda_2^2+2\lambda_1^2
\lambda_2^2)} {(\lambda^2+\lambda_1^2+\lambda_2^2-2\lambda \lambda_1 \lambda_2-1)^2},
\label{eq:30}
\end{eqnarray}
where the radial coordinates ${\lambda\equiv \cos\tilde{\theta}, \lambda_{1,2}
\equiv \cosh\tilde{\theta}_{1,2}}$ and the zero-mode action
\begin{eqnarray}
S&=&\frac{\pi}{2\Delta}[D_q(\Delta\theta)^2(2\lambda_1^2\lambda_2^2-\lambda^2-
\lambda_1^2-\lambda_2^2+1)\nonumber\\
&&\quad\quad + 2i\omega(\lambda-\lambda_1\lambda_2)].
\label{eq:31}
\end{eqnarray}
Substituting Eq.~(\ref{eq:30}) into Eq.~(\ref{eq:16}) we obtain Eq.~(\ref{eq:7}).

\end{appendix}


\begin{thebibliography}{99}
\bibitem{note1}        This symmetry is unimportant for irrational $\tilde h/(4\pi)$.
\bibitem{note2}        Exactly speaking, $E(t)$ such defined is the second moment of the wavepacket,
                       and thus characterizes the wavepacket spreading. It gives the genuine kinetic
                       energy only for quadratic $H_0$. Here we follow the convention to call it
                       kinetic energy.
\end{thebibliography}
\end{document}